\documentclass[a4paper,12pt,aip,jap,superscriptaddress]{revtex4-1}
\usepackage{graphicx}
\usepackage{booktabs}
\usepackage{multirow}
\usepackage{enumitem} 

\begin{document}

\title{Bimodal substrate biasing to control $\gamma$-Al$_2$O$_3$ deposition during reactive magnetron sputtering}

\author{Marina Prenzel} \email{marina.prenzel@rub.de}
\author{Annika Kortmann}
\author{Adrian Stein}
\author{Achim von Keudell}

\affiliation{Research Group Reactive Plasmas, Ruhr-Universit\"at Bochum, D-44801 Bochum,
Germany}

\author{Farwah Nahif}
\author{Jochen M. Schneider}

\affiliation{Materials Chemistry, RWTH Aachen University, D-52074 Aachen, Germany}

\begin{abstract}
Al$_2$O$_3$ thin films have been deposited at
substrate temperatures between 500$^\circ{}$C to 600$^\circ{}$C by
reactive magnetron sputtering using an additional arbitrary
substrate bias to tailor the energy distribution of the incident
ions. The films were characterized by X-ray diffraction (XRD) and
Fourier transform infrared spectroscopy (FTIR). The 
film structure being amorphous, nanocrystalline, or crystalline
was correlated with characteristic ion energy distributions.
The evolving crystalline structure is connected with different
levels of displacements per atom (dpa) in the growing film
as being derived from TRIM simulations. The boundary between
the formation of crystalline films and amorphous or
nanocrystalline films  was at 0.9 dpa for a substrate
temperature of 500$^\circ{}$C. This threshold shifts to 0.6 dpa
for films grown at 550$^\circ{}$C.

\end{abstract}

\maketitle

\section{Introduction}

Reactive magnetron sputtering (RMS) is a prominent technique to
deposit many thin film materials as for examples oxides and
nitrides using metal targets and the addition of oxygen and/or
nitrogen as a reactive component to the argon plasma gas
\cite{Koski1999,Clarke1993,Wallin2008b,Hoetzsch1996}.
The film stoichiometry and its structure can be adjusted by
controlling the ion-to-neutral ratio in the film forming growth
flux and the energy of the incident ions given by the ion energy
distribution function (IED). Incident ions may enhance the adatom
mobility and promote thereby crystallinity and/or a certain
crystalline orientation \cite{Wallin2008b,Rosen2005a,Rosen2005b}.
This is often quantified by the
energy per deposited atom $\langle E\rangle$. This energy depends
on the ion energy $E_{ions}$ and on the ion-to-neutral ratio
$J_{ions}$/$J_{growth}$ in the growth flux with $\langle
E\rangle$\,=\,$E_{ions}\cdot j_{ions}$/$j_{growth}$. Here, $j_{ions}$ is
the flux of incident ions and $j_{growth}$ the total flux of
incorporated atoms in the film.

Under variation of the average parameter $\langle E\rangle$
and the ion to neutral ratio $j_{ions}/j_{growth}$, it was
shown by Adibe {\it et al.} \cite{Adibi1993} and Petrov
{\it et al.} \cite{Petrov1993} that the average energy per
incorporated atom $\langle E\rangle$ is no universal parameter
for the formation of titanium nitride (Ti$_{0.5}$Al$_{0.5}$N).
However, in addition to Musil \textit{et al.} \cite{Musil1990},
the average energy per incorporated atom is suggested to
define the formation of crystallinity by several authors
\cite{Poulek1989,Poulek1991,Grigorov1991}.

Recently, we devised an
experiment to very accurately control the growth flux and the
energy distribution of the incident ions \cite{Prenzel2013} by
keeping the average energy per deposited atom $\langle E\rangle$,
the average energy of the incident ions $\langle E_{ions}
\rangle$, the total ion flux $j_{ions}$ and the ion-to-neutral
ratio $j_{ions}/j_{growth}$ constant, but changing {\it only} the
ion energy distribution (IED), for details see \cite{Prenzel2013}.

We applied this concept to RMS of Al$_2$O$_3$ films as a prominent
material with applications ranging from microelectronics, wear
resistant coatings to catalytic surfaces \cite{Hoetzsch1996}. The
most common phases of Al$_2$O$_3$ are the $\gamma$- and
$\alpha$-phase. $\alpha$-Al$_2$O$_3$ with its hexagonal closed
package (hcp) \cite{Gautier1994} structure is often used as hard
coating on machining tools.

In 2005 Rosen \textit{et al.} have reviewed phase formation data reported
for vapor phase deposited alumina \cite{Rosen2005b} and
summarize that the majority of authors observe that the crystalline
growth temperature appears to be reduced as the mobility of surface
species through energetic ion bombardment is increased. In 2010 Jiang
\textit{et al.} report an alpha alumina formation temperature of
560$^{\circ}$C by utilizing large ion fluxes during PACVD
\cite{Jiang2010} as discharge power densities of 19\,Wcm$^{-2}$
resulted in an increase in the energy and the flux of the bombarding
species towards the growing film, as well as in a more efficient
precursor dissociation \cite{Jiang2010}. In the same year, Sarakinos
\textit{et al.} reported an alpha alumina formation temperature of 720$^{\circ}$C
by cathodic arc deposition \cite{Sarakinos2010d} for substrate bias
potentials between -40\,V and -200\,V. Based on ab initio molecular
dynamics calculations \cite{Music2011a} subplantation of the
impinging Al is identified to cause significantly larger irradiation
damage and hence larger mobility in the gamma alumina as compared
to alpha alumina. Consequently, the enhanced mobility results in
the growth of the alpha phases at the expense of the gamma phase.

From the above discussion it can be learned that in addition to
the well-established mechanism of ion bombardment mediated surface
diffusion the previously overlooked subplantation mechanism
\cite{Sarakinos2010d,Music2011a} was suggested to be relevant for
the formation of crystalline alumina thin films.

Previously \cite{Prenzel2013}, we monitored the transition from
X-ray amorphous to $\gamma$-alumina to assess the influence of the ion
energy distribution on thin film growth, indicating that typically
one displacement per incorporated atom (dpa) is necessary for
that transition to occur. In this paper, we expand the data set
and measured a comprehensive set of XRD and FTIR data to also
identify nanocrystalline samples, which appear as being crystalline
in the FTIR measurement although they are still X-ray amorphous.
Based on this large data set, the hypothesis of characteristic dpa
levels to allow for a certain structural transitions in the films
can be more thoroughly tested. Formation of amorphous films,
of nanocrystalline films or of X-ray $\gamma$-crystalline
Al$_2$O$_3$ films was observed by FTIR and XRD. It has to be
emphasized that our parameter interval is restricted to only the
variation of the ion energy distribution function of the incoming
ions and the substrate temperature. Therefore, it is possible
to isolate the effect of the transfer of kinetic energy on the
film growth.\clearpage

\section{Experimental methods}
\subsection{Film deposition}

Thin aluminum oxide films were deposited using a dual frequency
magnetron sputter experiment employing 13.56\,MHz and 71\,MHz for
plasma generation, as described in detail in \cite{Prenzel2013}.
The discharge was operated at 0.1\,Pa at a constant argon flow
rate of 9\,sccm. The base pressure in the deposition chamber was
2$\cdot$10$^{-5}$\,Pa. Based on \cite{Rosen2006,Prenzel2013}
this is expected to lead to H incorporation of $<$2at.\%. The
incorporated H atoms from water in the residual gas \cite{Schneider1999}
is small enough that the effect of incorporated water in the
films can be neglected. 
A feedback loop regulated the oxygen flow into
the chamber to avoid target poisoning by monitoring a constant
intensity of the Al I emission line at 396.2\,nm using a narrow
band pass filter and a photomultiplier. The adjusted oxygen partial
pressure in the deposition chamber was determined as 8.6$\cdot$10$^{-3}$\,Pa
in average. Thereby, stoichiometric Al$_2$O$_3$ coatings were
prepared as being verified by ex-situ X-ray Photoelectron Spectroscopy (XPS).

The distance between target and the p-doped Si(100) substrate was
50\,mm. The substrate temperature during the deposition process
was regulated to 500\,$^{\circ}$C, 550\,$^{\circ}$C and
600\,$^{\circ}$C, respectively. The substrate temperature was
directly measured by analyzing the temperature dependent
refraction index of silicon at 632.8\,nm by ellipsometry
\cite{Kroesen1991}. This is non intrusive and measures directly
the surface temperature. Any thermocouple at the substrate holder
my read a different temperature due to improper thermal contacts.
Pyrometry remain ambiguous because the emissivity of the
coated silicon wafer is not well defined and silicon becomes
transparent in the infrared wavelength range at high temperatures.

\subsection{Substrate biasing and ion energy distribution}

The substrate electrode was intentionally biased to tailor the ion
energy distribution function (IEDF) of the incident ions.
Rectangular waveforms were generated by a waveform generator and
amplified using a broadband amplifier. A coupling capacitor was
used to connect the biasing signal to the substrate electrode.

The kinetic energy of the impinging ions is controlled as follows:
a rectangular biasing signal can be divided into an on-time
($\tau_{on}$) and an off-time ($\tau_{off}$). The frequency of the
pulsing $f$ is given as $f=1/(\tau_{on}+\tau_{off})$. In our
experiments, the on-time $\tau_{on}$ was fixed to a value of 500\,ns and a
change in frequency $f$ of the applied biasing signal was realized
by changing the off-time $\tau_{off}$ only. Thereby, the fluence
($=j_{ions}\tau_{on}$) of energetic ions during growth remains
identical in all experiments. The maximum ion energy
$E_{ions,max}$, in case of collisionless sheaths, corresponds to
the voltage drop between the biasing signal $U_{max}$ and the
plasma potential. Consequently, $E_{ions,max}$ was used as
parameter to uniquely characterize the ion bombardment during film
growth.

\begin{figure}[h]
    \centering
        \includegraphics[width=13cm]{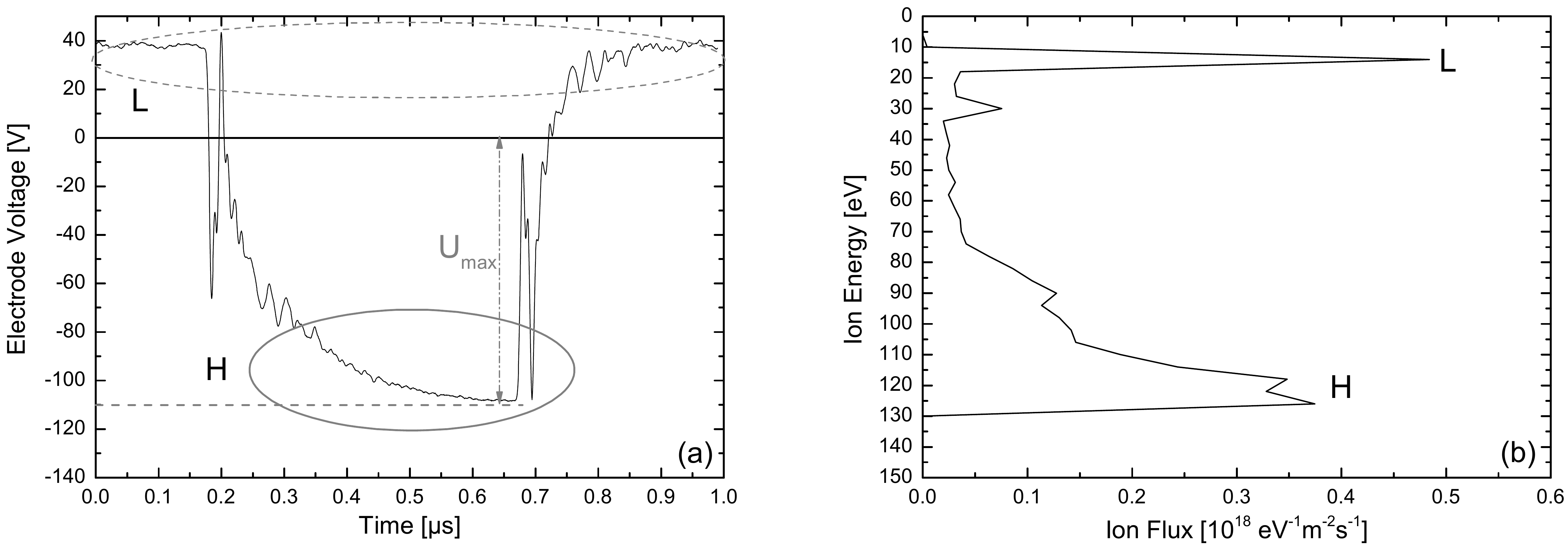}
    \caption{(a) Rectangular biasing signal at the substrate electrode at 1.01\,MHz, as measured by the oscilloscope.
    (b) simulated ion energy distribution function IEDF for the given bias signal by the model from Shihab et al. \cite{Shihab2012}.
    L and H indicate the low and high energy part of the IEDF, respectively.}
    \label{fig:Bias}
\end{figure}

A typical bias signal for a frequency $f$=1.01\,MHz and a
maximum voltage of $U_{max}=-110\,V$ is shown in Figure
\ref{fig:Bias} (a) as being measured at the substrate electrode by
an oscilloscope. The signal is not perfectly rectangular due to
the low pass filtering effect of the coupling capacitor. The
resulting IEDF was simulated from the voltage signal at the
electrode by a sheath model, as described by Shihab \textit{et
al.} \cite{Shihab2012}. The resulting IEDF is shown in Figure
\ref{fig:Bias} (b) as ion flux per ion energy interval versus ion
energy.

Two prominent peaks (regions H and L in Figure \ref{fig:Bias}) in
the IEDF can be identified. The maximum ion energy $E_{ions,max}$
within the IEDF in Figure \ref{fig:Bias} (b) corresponds to the
maximum bias voltage applied to the substrate electrode
$U_{max}$ (region H). During this time period, ions are
accelerated by the voltage drop between plasma potential (18\,V)
and maximum bias voltage to the substrate electrode
($U_{max}$). This results in a maximum ion energy of
$E_{ions,max}=18\,eV-U_{max}\cdot e$. In addition, a low energetic
peak (region L) originates from ions accelerated during the
off-time  $\tau_{off}$ with an ion energy corresponding to the
voltage drop between plasma potential and floating potential.

Frequencies $f$ between 0.80\,MHz and 1.60\,MHz were applied to
regulate the ratio between the ion flux at high vs. low ion
energies. The ion energy of the high energetic peak within the
IEDF was adjusted to a desired ion energy by a defined tailoring of
the maximum bias voltage $U_{max}$. Therefore, any ratio
between the two peaks within the IEDF and any position of the high
energetic peak can be reached by tailoring frequency $f$ and
maximum bias voltage $U_{max}$.

The energy impact during film growth is usually defined by the mean
energy per incorporated atom $\langle E\rangle$ depending on the
maximum ion energy of the ions $E_{ions,max}$, the total growth
flux $j_{growth}$, the ion flux $j_{ions}$ and the duty cycle
$d.c.= \tau_{on}\cdot f$ according to:

        \begin{equation}
        \langle E \rangle=E_{ions,max}\cdot\frac{j_{ions}}{j_{growth}}\cdot\tau_{on}\cdot f \label{eq:meanenergy}
        \end{equation}

The growth flux $j_{growth}$ is defined by the incorporated flux
of aluminum and oxygen atoms during Al$_2$O$_3$ film formation.
This is deduced from the total film thickness, as measured ex situ
by a profilometer, divided by the overall deposition time. The
growth rate $g$ expressed in nm s$^{-1}$ is converted into the
growth flux in cm$^{-2}$s$^{-1}$ using the density of the film
$\rho$ and the average mass $M$ of Al$_2$O$_3$ via
$j_{growth}=\frac{g\cdot \rho}{M}$. A residual thickness
inhomogeneity of 4\% is observed which converts into an error
of the growth  flux of 5\%.

Because XRD analysis of our films exhibits good crystalline
quality of $\gamma$-alumina (see below), the film density value from literature
of $\rho$=3.66\,g$\cdot$cm$^{-3}$ as being reported by Levin \textit{et al.}
\cite{Brandon1998} seems to have reasonable good agreement
with our films and is taken for the further discussion.

Finally, the ion flux to the substrate surface is required when
determining the mean ion energy per incorporated atom $\langle E
\rangle$. The ion flux $j_{ions}$ was measured using a retarding field analyzer
within a previous work \cite{Prenzel2012}. It was determined as
$j_{ions}$=13.5$\,\cdot\,10^{18}$\,m$^{-2}$s$^{-1}$.

\subsection{Thin film analysis}

Phase formation was studied by XRD with a Bruker D8 General Area Diffraction System
(GADDS) on the deposited Al$_{2}$O$_{3}$ thin films. The incident
angle of the beam was 15$^{\circ}$ and the analyzed 2$\Theta$ angle
range was 20$^{\circ}$ to 75$^{\circ}$. The applied
voltage and current settings were 40\,kV and 40\,mA, respectively.
Three different peaks which can be associated to the $\gamma$
phase of Al$_{2}$O$_{3}$ are identified in the diffraction
patterns of our samples. The (311) direction can be
found within the XRD pattern at an angle of 37.30$^{\circ}$.
Moreover, peaks at 45.86$^{\circ}$ and 67.03$^{\circ}$ are
identified as (400) and (440) orientations, respectively. Peak
positions agree with the JCDPS file number 10-0425
for $\gamma$-Al$_2$O$_3$ at 45.9$^{\circ}$ and 67.0$^{\circ}$.

Further analysis of the Al$_2$O$_3$ samples was realized by
ex situ FTIR transmission measurements using a Bruker IFS 66/S
spectrometer. A polarizer was placed in front of the sample, so
that only s polarized light reached the sample. The angle of
incidence normal to the surface was 60$^{\circ}$. Measurements
were performed in the wavenumber range between 400\,cm$^{-1}$ and
6,000\,cm$^{-1}$. Background spectra of non-coated silicon wafers
were used.

Br{\"u}esch \textit{et al.} \cite{Bruesch1984} investigated FTIR
spectra of amorphous and $\gamma$-aluminum oxide. The evolution of
a sharp peak (or dip within transmission spectra) at
950\,cm$^{-1}$ is characteristic for $\gamma$-Al$_{2}$O$_{3}$.
Further broad oscillations at lower wavenumbers being characterized by
Chu \textit{et al.} \cite{Chu1988a} at 357\,cm$^{-1}$,
536\,cm$^{-1}$ and 744\,cm$^{-1}$ were identified.

XRD and FTIR assess the crystallinity of the samples on distinct
length scales: (i) in XRD, the coherent scattering of the incident
X-radiation from a crystal leads to pronounced peaks in the XRD
diffractogram. In case of nanocrystalline samples, significant
line broadening occurs which complicates the evaluation of X-ray
profiles. As a consequence, the distinction between crystalline
and amorphous samples depends on the employed diagnostic method and 
nanocrystalline samples may not be detected by XRD. (ii) in FTIR,
the signal originates from the absorption of single Al-O bonds at
a frequency depending on the configuration of the next neighbors.
If nanocrystallites are formed, characteristic LO or TO phonon
peaks may appear in the infrared spectrum, whereas the coherent
overlap of the scattered light in XRD by crystalline region and
amorphous sites may still show X-ray amorphous diffraction patterns.

\begin{figure}[ht]
    \centering
        \includegraphics[width=12cm]{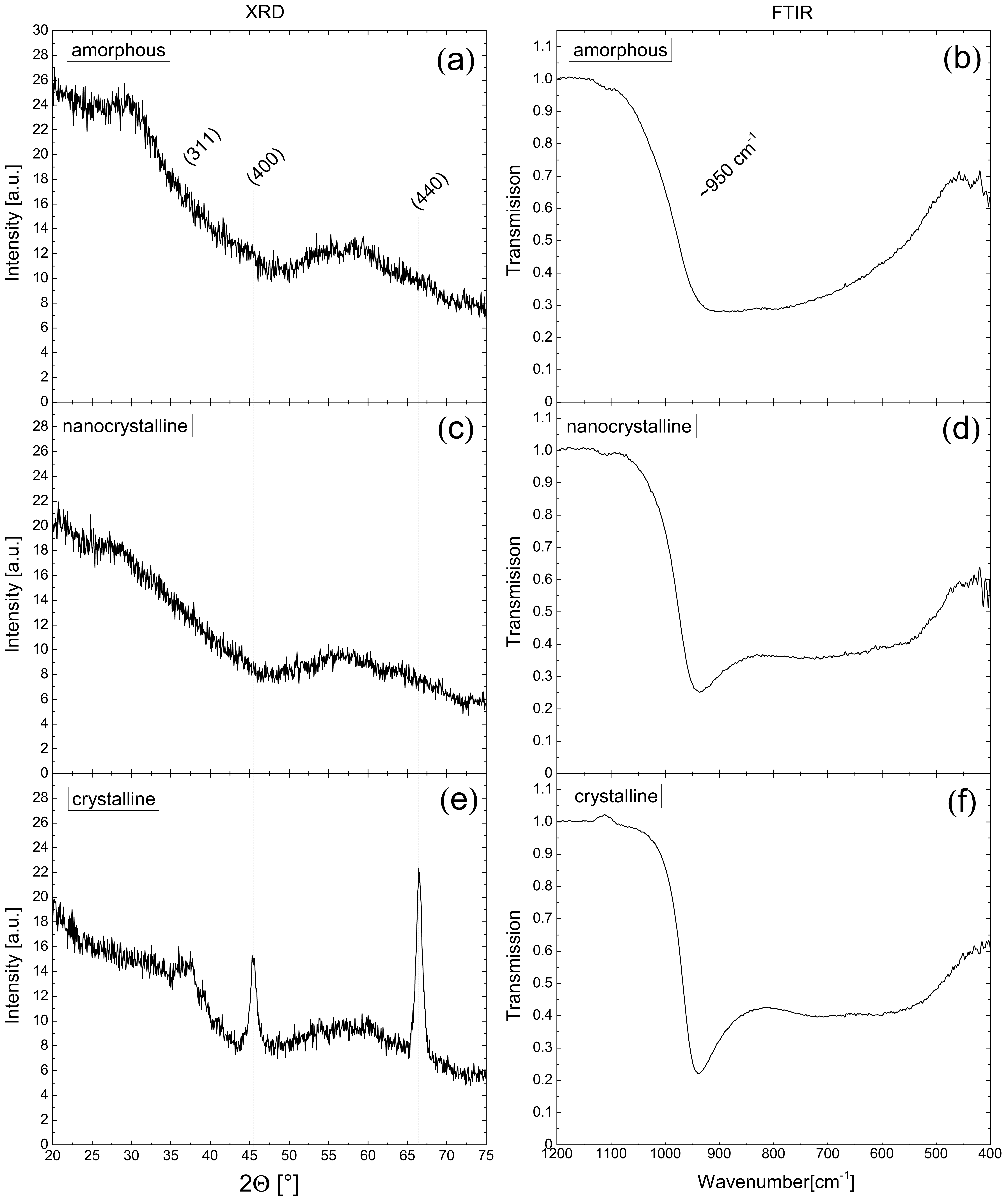}
    \caption{Typical XRD- and FTIR-spectra for amorphous samples (a and b),
    for nanocrystalline samples (c and d),
    and for crystalline samples (e and f).}
    \label{fig:XRDFTIR}
\end{figure}

Based on the different measurement principles of XRD and FTIR, the
transition from amorphous, nanocrystalline, to crystalline samples
can be assessed, as illustrated in Figure \ref{fig:XRDFTIR} for
films deposited at different deposition conditions, as described
below: (i) Figure \ref{fig:XRDFTIR} show the spectra for amorphous
films, since no distinct peaks in the XRD diffractogram (a)
 and no sharp absorption for TO phonon at approximately 950\,cm$^{-1}$
is found (b) ; (ii) Figure
\ref{fig:XRDFTIR}  shows an XRD diffractogram (c) for a nanocrystalline
film, which appears XRD-amorphous. Infrared absorption at a wavenumber of
950\,cm$^{-1}$ reflects the formation of nanocrystallites, which
is shown in Figure \ref{fig:XRDFTIR} (d); (iii) Figure
\ref{fig:XRDFTIR} shows an XRD diffractogram (e) for a crystalline film
with sharp peaks in the XRD diffractogram at position characteristic to
$\gamma$-alumina. In addition to the XRD signature, pronounced
absorptions for TO phonon become also visible in the IR spectrum
(f).

It is important to note that the identification of the
nanocrystalline phase from a comparison of the XRD and FTIR data
remains ambiguous for very few cases, because the sharpness of the
TO-phonon peak in FTIR is not as distinct as the analysis of the
diffraction peaks in the XRD diffractogram. Those samples are marked in
the presented data sets.\clearpage

\section{Results and Discussion}

\begin{figure}[ht]
    \centering
        \includegraphics[width=8cm]{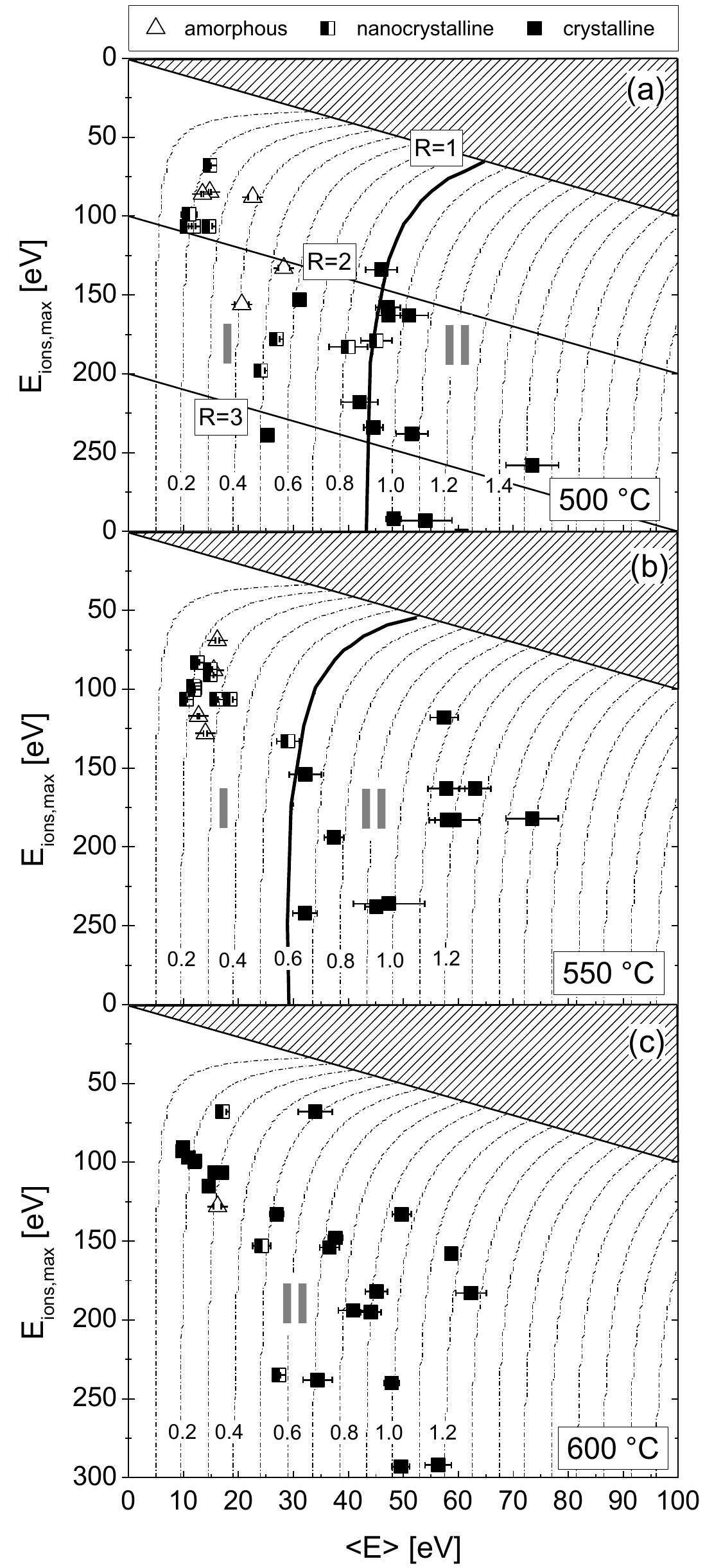}
    \caption{Al$_2$O$_3$ film structure as identified from the XRD- and FTIR spectra being amorphous (open triangles),
    nanocrystalline (half filled squares), or crystalline (filled
    squares) in dependence on the average energy per incorporated atom $\langle E \rangle$ and the maximum ion energy $E_{ions,max}$.
    The shaded areas exclude unreasonable combinations of $\langle E \rangle$ and $E_{ions,max}$. The substrate temperature during
    deposition was set at 500\,$^{\circ}$C (a), at 550\,$^{\circ}$C (b), or at 600\,$^{\circ}$C.
    The contour lines show the displacements per atom (dpa), as calculated from the displacements
    per ions (dpi) simulated by TRIM and the ion-to-neutral ratio in the growth flux.
    The measured film structures are separated into two regions I and II,
    characterized by different dpa levels.
    The ion flux to total growth flux ratio R is shown as straight lines in (a).}
    \label{fig:al2o3}
\end{figure}

A large set of 73 samples was prepared in the substrate
temperature range between 500\,$^{\circ}$ and 600\,$^{\circ}$C at a
pressure of 0.1\,Pa. The IEDF was tailored using frequencies $f$ in
the range between 0.80\,MHz and 1.60\,MHz for the pulsed bias and
maximum bias voltages between -25\,V to -280\,V. The range of
operating parameters and the resulting average energies $\langle E
\rangle$ are listed  in Table \ref{tab}. The samples were analyzed
ex situ by XRD and FTIR with respect to their structure. The deposition is
characterized by the maximum ion energy $E_{ions,max}$ and the
average energy per incorporated atom $\langle E \rangle$. Figure
\ref{fig:al2o3} shows the results for the amorphous samples (open
triangle), the nanocrystalline samples (half filled squares), and
the crystalline samples (solid squares) for three different
substrate temperatures of 500\,$^\circ$C (Figure \ref{fig:al2o3}a),
of 550\,$^\circ$C (Figure \ref{fig:al2o3}b), and of 600\,$^\circ$C
(Figure \ref{fig:al2o3}c). The dashed areas in Figure \ref{fig:al2o3}
indicate deposition parameters, which are not reasonable, because
the average energy $\langle E \rangle$ cannot be larger than the
maximum energy $E_{ions,max}$.

Finally, the correlation of the deposition parameters with the
structure of the deposited films may be affected by nucleation
phenomena.

\begin{table}[b]
\begin{centering}
\begin{tabular}{|l|c|}
\hline Deposition/&Adjusted values\tabularnewline Biasing
parameter&\tabularnewline \hline
Temperature T &500\,$^{\circ}$C, 550\,$^{\circ}$C, 600\,$^{\circ}$C\\[4pt]
\multirow{3}{*}{Biasing frequency f} & 0.80\,MHz, 1.01\,MHz,\\
&1.20\,MHz, 1.40\,MHz,\\
&1.60\,MHz\\
Mean energy per incor-&\multirow{2}{*}{10\,eV...30\,eV...60\,eV}\\
porated atom $<E>$ &\\[4pt]
Maximum Biasing&\multirow{2}{*}{-25\,V...-280\,V}\\
voltage $U_{max}$&\\[4pt]
Maximum ion&\multirow{2}{*}{7\,eV...262\,eV}\\
energy $E_{ions,max}$&\tabularnewline \hline
\end{tabular}
\caption{Deposition parameters for Al$_2$O$_3$ growth by reactive
magnetron sputtering with an additional substrate
bias.}\label{tab}
\end{centering}
\end{table}

The data in Figure \ref{fig:al2o3} indicate that crystalline films
are usually obtained if the average energy $\langle E \rangle$ and
the maximum energy $E_{ions,max}$ are above a certain threshold.
This becomes more critical at lower substrate temperatures, where
also amorphous films are observed at low $\langle E \rangle$ and
$E_{ions,max}$. The films at intermediate values for $\langle E
\rangle$ and $E_{ions,max}$ show nanocrystalline behavior. At a
temperature of 600\,$^{\circ}$C, aluminum oxide is deposited in
the $\gamma$ phase even for very low ion bombardment and only very
few samples remain amorphous.

This rough analysis already illustrates that an increasing energy
input during film growth induces a transition of the film structure
from an amorphous to a nanocrystalline and finally to a
crystalline structure. This is in agreement with the current
general understanding of energetic film deposition.

The ion-induced formation of a nanometer size crystal or the
$\gamma$ phase could be induced by displacement events within
a collision cascade. These displacements generate mobility and
may turn enable the the formation of a crystalline structure.
This very general picture can be tested with
our data by comparing the samples, as plotted in Figure
\ref{fig:al2o3}, with TRIM simulations \cite{TRIM2012} to
calculate the displacements per incorporated atoms in the growing
films.

\begin{figure}[h]
    \centering
        \includegraphics[width=9cm]{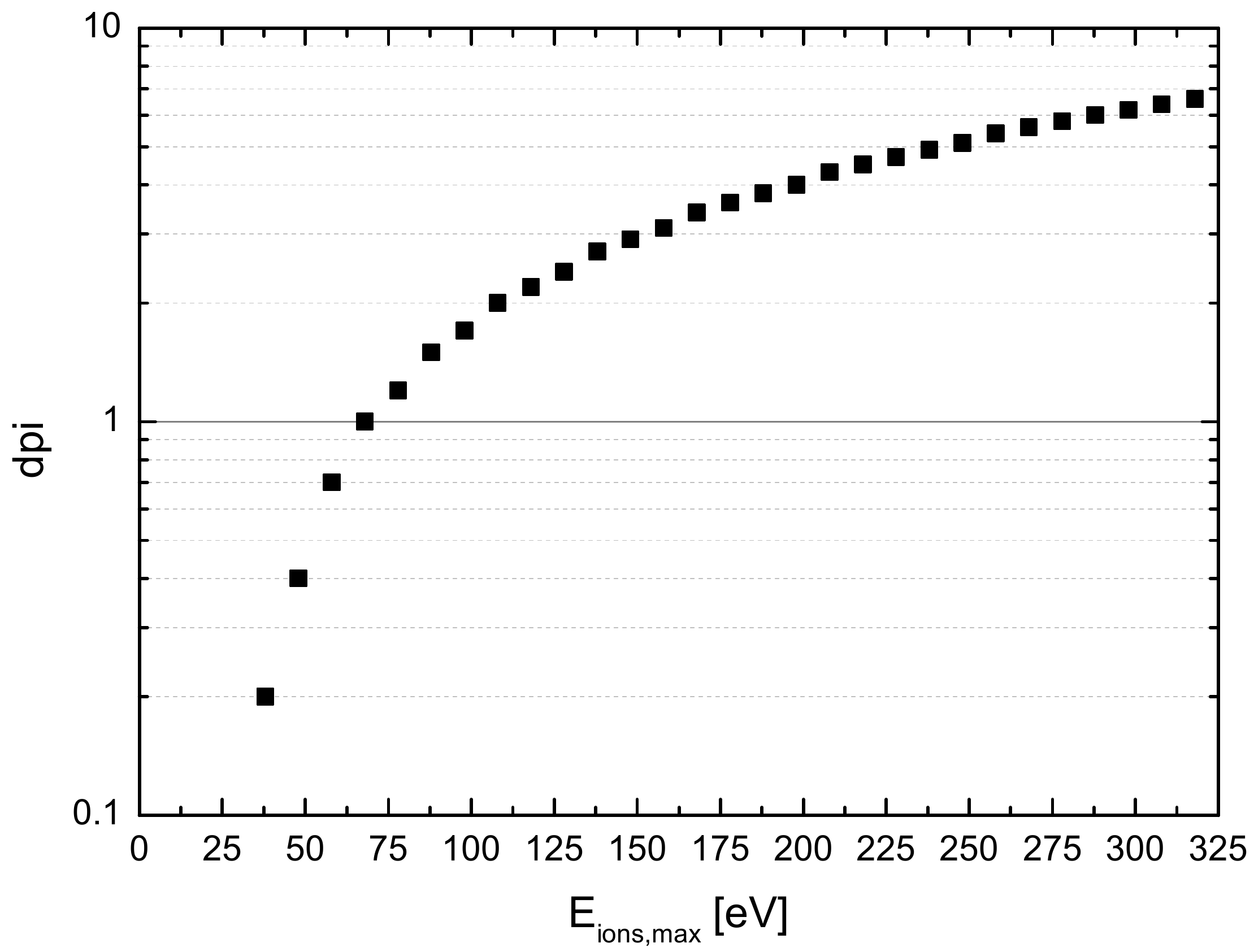}
    \caption{TRIM calculation for displacement per ion (dpi) for Ar ions impinging onto an Al$_2$O$_3$ surface.}
    \label{fig:dpi}
\end{figure}

TRIM simulations were performed for argon ions, which initiate a
collision cascade in an stoichiometric Al$_2$O$_3$ film. Surface
binding energies for Al atoms and O were assumed as 3.36\,eV and
2.00\,eV, respectively. A density of 3.66\,g\,cm$^{-3}$ was
assumed \cite{Brandon1998} and mono energetic argon ions with
energies between 28\,eV and 308\,eV. This corresponds to maximum
voltages at the substrate electrode $U_{max}$ of -10\,V to
-300\,V, respectively. The displacement per ion (dpi) for the
different ion energies was calculated and plotted versus ion
energy in Figure \ref{fig:dpi}. No distinction between displacing
aluminum or oxygen is made. A threshold energy of 32\,eV for
aluminum oxide is necessary to induce a displacement event
within the Al$_2$O$_3$ film. Ion bombardment by a rectangular
bias signal establishes two distinct peaks with
different ion energies impinging onto the substrate surface.
Ions with low energies with 18\,eV are below the threshold to
initiate a displacement within the aluminum oxide film.
Only ions with ion energies above the critical value of 32\,eV
have enough energy to initiate a displacement within the film.
The low energetic ions are accelerated within the off period of
the bias signal, whereas high energetic ions are produced
during the on-time of the signal. Therefore, only ions from
the high energetic peak within the IEDF account to the dpi within
the Al$_2$O$_3$ films.\clearpage

The displacements per ion (dpi) are converted into displacements
per atom (dpa) during film growth by multiplying it with the ratio
between energetic ion fluence per puls $j_{ions} \tau_{on}$ and
the growth fluence $j_{growth} (\tau_{on}+\tau_{off})$:

\begin{equation}
dpa=dpi\cdot\frac{j_{ions}}{j_{growth}}
\frac{t_{on}}{t_{on}+t_{off}}=dpi\cdot\frac{j_{ions}}{j_{growth}}
t_{on}\cdot f\label{eq:dpa}
        \end{equation}

Equation \ref{eq:meanenergy} can be combined with eq. \ref{eq:dpa}
yielding a dependence of the dpa level on the control parameters
$E_{ions,max}$ and $\langle E \rangle$:

\begin{equation}
dpa=dpi\cdot\frac{\langle E\rangle}{E_{ions,max}}  \label{eq:dpa2}
\end{equation}\clearpage

\thispagestyle{empty}
\clearpage

The resulting dpa levels are then plotted as contour lines in Figure
\ref{fig:al2o3}, separating the samples into two regions
(indicated as I and II in Figure \ref{fig:al2o3}) for
different substrate temperatures:

\begin{itemize}
\item \textbf{Deposition at 500$^{\circ}$C}: in region I (dpa $<$ 0.9),
most films are amorphous or nanocrystalline, while in region II
(dpa $>$ 0.9) all films are crystalline. The threshold may also
be aligned to a certain mean energy per deposited atom
$\langle E \rangle$, which might be detected at 43\,eV.

The observation of different dpa threshold values being necessary
for a structural transformation from amorphous to nanocrystalline
or from nanocrystalline to amorphous, respectively, is consistent
with the current understanding of energetic film formation: a
certain energy input is required to allow for the formation of a
specific phase of the material.

The separation of the samples into different regions depending on
the dpa level is similar to a criterion solely based on the
average energy per deposited atoms for ion energies above 150\,eV.
This is illustrated by the contour lines being almost parallel to
the y-axis due to the linear relationship between dpi and ion
energy at high energies (see Figure \ref{fig:dpi}).

The separation of the samples into different regions may be
regarded in more detail. Petrov \textit{et al}. \cite{Adibi1993} made an
observation indicating that besides an average energy, the
absolute energy has to be at least high enough for the incident
ion to penetrate into the material and to displace atoms. This
criterion, however, is already the basis of the dpa contour plots.
In our case, we still see structural changes although the
dpa level remains the same if we follow individual contour lines
at the border between the regions. The only remaining difference for
those samples is the fact that the dissipated energy is
distributed among several species for $E_{ions,max}$ $>$ 200\,eV,
but dissipated only by a few atoms for $E_{ions,max}$ $<$ 150\,eV.
The number of atoms that dissipate the ion energies are given by
the flux ratios, which are additionally plotted in Figure
\ref{fig:al2o3} (a).

\item \textbf{Deposition at 550$^{\circ}$C}: in region I
(dpa $<$ 0.6) most films are amorphous or nanocrystalline,
while above a dpa value of 0.6 all films become crystalline.
One can clearly see that the boundaries between amorphous
or nanocrystalline films to crystalline films shifts to
lower dpa levels. This is consistent with the current
understanding of film formation that less ion-induced mobility
is required at higher substrate temperatures.

\item \textbf{Deposition at 600$^{\circ}$C}: all films become
crystalline, irrespective of the dpa level. Apparently,
the substrate temperature is high enough so that even
very small dpa levels are already sufficient to induce
the formation of crystalline films. A clear transition
between region II and I cannot be identified anymore in our
data set.
\end{itemize}

A close inspection of Figure \ref{fig:al2o3} shows that very
few samples are not located consistently in the respective
region - crystalline films in region I instead of II, or
amorphous/ nanocrystalline films in region II instead of I. This
deviation may be induced by two effects, the variation in
the nucleation process or the variation in the grain size
distribution:
\begin{enumerate}[label=({\alph*})]
\item In nucleation a delicate competition
between amorphous and crystalline phases can occur, where
small deviations in the initial condition such as substrate
morphology, contamination etc. may lead to growth of
completely different film structures.

\item The distinction
between crystalline and amorphous films by XRD depends very
sensitive on the grain size distribution. For small grains
or low grain size density, a crystalline phase might easily
be undetected by XRD. However, it should be emphasized that
the vast majority of samples can consistently be located in
region I and II, respectively.
\end{enumerate}

The mapping of the film structure on the parameters of the energy
input during film growth expressed in $E_{ions,max}$ and $\langle
E \rangle$ and its comparison to the dpa levels that are induced
during film growth shows good agreement. One may extrapolate this
scaling to Al$_2$O$_3$ deposition in general, to predict the film
structure in the plasma based on the operating parameters of the
system. The description of the IEDF with a maximum ion energy is
only reasonable for rectangular bias. The underlying important
quantity however is the dpa level in the film, which can be
uniquely calculated using TRIM calculations for any ion energy
distribution function. The extrapolation of the proposed scaling
approach for other bias waveforms is currently under way.\clearpage

\section{Conclusion}
Al$_2$O$_3$ thin films have been deposited at substrate
temperatures between 500$^\circ{}$C and 600$^\circ{}$C by reactive
magnetron sputtering using an additional arbitrary substrate bias
to tailor the energy distribution of the incident ions. The formation
of crystalline films as opposed to amorphous or nanocrystalline films
depends on a critical threshold of 0.9 displacement per incorporated
atom at 500$^\circ{}$C substrate temperature. This threshold
shifts to 0.6 dpa with increasing substrate temperature to
550$^\circ{}$C. One can conclude that for fixed neutral to ion fluxes
to the film surface, the dpa value is a predictor for the formation
of crystalline Al$_2$O$_3$ films. The driving mechanism of the
formation of crystalline structures is the enhanced mobility
of surface atoms due to Ar ion bombardment. As the dpa value
increases, the formation temperature of crystalline Al$_2$O$_3$
thin films is decreased.

\section{Acknowledgment}

The authors would like to thank Norbert Grabkowski for his
technical support within the experimental setup.

This project is supported by DFG (German
Research Foundation) within the framework of the Special Research
Field SFB-TR 87 and the Research Department `Plasmas with Complex
Interactions' at Ruhr-Universität Bochum.

Finally, the authors like to thank M. Shihab and R.P. Brinkmann
for their support in modeling the ion energy distribution
functions for a given bias signal at the electrodes.\clearpage

\bibliography{library}

\begin{thebibliography}{26}
\expandafter\ifx\csname natexlab\endcsname\relax\def\natexlab#1{#1}\fi
\expandafter\ifx\csname bibnamefont\endcsname\relax
  \def\bibnamefont#1{#1}\fi
\expandafter\ifx\csname bibfnamefont\endcsname\relax
  \def\bibfnamefont#1{#1}\fi
\expandafter\ifx\csname citenamefont\endcsname\relax
  \def\citenamefont#1{#1}\fi
\expandafter\ifx\csname url\endcsname\relax
  \def\url#1{\texttt{#1}}\fi
\expandafter\ifx\csname urlprefix\endcsname\relax\def\urlprefix{URL }\fi
\providecommand{\bibinfo}[2]{#2}
\providecommand{\eprint}[2][]{\url{#2}}

\bibitem[{\citenamefont{Koski et~al.}(1999)\citenamefont{Koski, H\"{o}ls\"{a},
  and Juliet}}]{Koski1999}
\bibinfo{author}{\bibfnamefont{K.}~\bibnamefont{Koski}},
  \bibinfo{author}{\bibfnamefont{J.}~\bibnamefont{H\"{o}ls\"{a}}},
  \bibnamefont{and} \bibinfo{author}{\bibfnamefont{P.}~\bibnamefont{Juliet}},
  \bibinfo{journal}{Thin Solid Films} \textbf{\bibinfo{volume}{339}},
  \bibinfo{pages}{240} (\bibinfo{year}{1999}), ISSN \bibinfo{issn}{00406090},
  \urlprefix\url{http://linkinghub.elsevier.com/retrieve/pii/S0040609098012322}.

\bibitem[{\citenamefont{Clarke}(1994)}]{Clarke1993}
\bibinfo{author}{\bibfnamefont{P.~J.} \bibnamefont{Clarke}},
  \bibinfo{journal}{Journal of Vacuum Science \& Technology A: Vacuum,
  Surfaces, and Films} \textbf{\bibinfo{volume}{12}}, \bibinfo{pages}{594}
  (\bibinfo{year}{1994}), ISSN \bibinfo{issn}{07342101},
  \urlprefix\url{http://link.aip.org/link/?JVA/12/594/1\&Agg=doi}.

\bibitem[{\citenamefont{Wallin et~al.}(2008)\citenamefont{Wallin, Selinder,
  Elfwing, and Helmersson}}]{Wallin2008b}
\bibinfo{author}{\bibfnamefont{E.}~\bibnamefont{Wallin}},
  \bibinfo{author}{\bibfnamefont{T.~I.} \bibnamefont{Selinder}},
  \bibinfo{author}{\bibfnamefont{M.}~\bibnamefont{Elfwing}}, \bibnamefont{and}
  \bibinfo{author}{\bibfnamefont{U.}~\bibnamefont{Helmersson}},
  \bibinfo{journal}{EPL (Europhysics Letters)} \textbf{\bibinfo{volume}{82}},
  \bibinfo{pages}{36002} (\bibinfo{year}{2008}), ISSN
  \bibinfo{issn}{0295-5075},
  \urlprefix\url{http://stacks.iop.org/0295-5075/82/i=3/a=36002?key=crossref.fada472bbee68e1835daeba98887bc9a}.

\bibitem[{\citenamefont{Zywitzki et~al.}(1996)\citenamefont{Zywitzki, Hoetzsch,
  Fietzke, and Goedicke}}]{Hoetzsch1996}
\bibinfo{author}{\bibfnamefont{O.}~\bibnamefont{Zywitzki}},
  \bibinfo{author}{\bibfnamefont{G.}~\bibnamefont{Hoetzsch}},
  \bibinfo{author}{\bibfnamefont{F.}~\bibnamefont{Fietzke}}, \bibnamefont{and}
  \bibinfo{author}{\bibfnamefont{K.}~\bibnamefont{Goedicke}},
  \bibinfo{journal}{Surface and Coatings Technology}
  \textbf{\bibinfo{volume}{82}}, \bibinfo{pages}{169} (\bibinfo{year}{1996}),
  ISSN \bibinfo{issn}{02578972},
  \urlprefix\url{http://linkinghub.elsevier.com/retrieve/pii/0257897295002707}.

\bibitem[{\citenamefont{Ros\'{e}n
  et~al.}(2005{\natexlab{a}})\citenamefont{Ros\'{e}n, Schneider, and
  Larsson}}]{Rosen2005a}
\bibinfo{author}{\bibfnamefont{J.}~\bibnamefont{Ros\'{e}n}},
  \bibinfo{author}{\bibfnamefont{J.~M.} \bibnamefont{Schneider}},
  \bibnamefont{and} \bibinfo{author}{\bibfnamefont{K.}~\bibnamefont{Larsson}},
  \bibinfo{journal}{Solid State Communications} \textbf{\bibinfo{volume}{135}},
  \bibinfo{pages}{90} (\bibinfo{year}{2005}{\natexlab{a}}), ISSN
  \bibinfo{issn}{00381098},
  \urlprefix\url{http://linkinghub.elsevier.com/retrieve/pii/S0038109805003078}.

\bibitem[{\citenamefont{Ros\'{e}n
  et~al.}(2005{\natexlab{b}})\citenamefont{Ros\'{e}n, Mr\'{a}z, Kreissig,
  Music, and Schneider}}]{Rosen2005b}
\bibinfo{author}{\bibfnamefont{J.}~\bibnamefont{Ros\'{e}n}},
  \bibinfo{author}{\bibfnamefont{S.}~\bibnamefont{Mr\'{a}z}},
  \bibinfo{author}{\bibfnamefont{U.}~\bibnamefont{Kreissig}},
  \bibinfo{author}{\bibfnamefont{D.}~\bibnamefont{Music}}, \bibnamefont{and}
  \bibinfo{author}{\bibfnamefont{J.~M.} \bibnamefont{Schneider}},
  \bibinfo{journal}{Plasma Chemistry and Plasma Processing}
  \textbf{\bibinfo{volume}{25}}, \bibinfo{pages}{303}
  (\bibinfo{year}{2005}{\natexlab{b}}), ISSN \bibinfo{issn}{0272-4324},
  \urlprefix\url{http://www.springerlink.com/index/10.1007/s11090-004-3130-y}.

\bibitem[{\citenamefont{Adibi et~al.}(1993)\citenamefont{Adibi, Petrov, Greene,
  Hultman, and Sundgren}}]{Adibi1993}
\bibinfo{author}{\bibfnamefont{F.}~\bibnamefont{Adibi}},
  \bibinfo{author}{\bibfnamefont{I.}~\bibnamefont{Petrov}},
  \bibinfo{author}{\bibfnamefont{J.~E.} \bibnamefont{Greene}},
  \bibinfo{author}{\bibfnamefont{L.}~\bibnamefont{Hultman}}, \bibnamefont{and}
  \bibinfo{author}{\bibfnamefont{J.-E.} \bibnamefont{Sundgren}},
  \bibinfo{journal}{Journal of Applied Physics} \textbf{\bibinfo{volume}{73}},
  \bibinfo{pages}{8580} (\bibinfo{year}{1993}), ISSN \bibinfo{issn}{00218979},
  \urlprefix\url{http://jap.aip.org/resource/1/japiau/v73/i12/p8580\_s1
  http://link.aip.org/link/JAPIAU/v73/i12/p8580/s1\&Agg=doi}.

\bibitem[{\citenamefont{Petrov et~al.}(1993)\citenamefont{Petrov, Adibi,
  Greene, Hultman, and Sundgren}}]{Petrov1993}
\bibinfo{author}{\bibfnamefont{I.}~\bibnamefont{Petrov}},
  \bibinfo{author}{\bibfnamefont{F.}~\bibnamefont{Adibi}},
  \bibinfo{author}{\bibfnamefont{J.~E.} \bibnamefont{Greene}},
  \bibinfo{author}{\bibfnamefont{L.}~\bibnamefont{Hultman}}, \bibnamefont{and}
  \bibinfo{author}{\bibfnamefont{J.-E.} \bibnamefont{Sundgren}},
  \bibinfo{journal}{Applied Physics Letters} \textbf{\bibinfo{volume}{63}},
  \bibinfo{pages}{36} (\bibinfo{year}{1993}), ISSN \bibinfo{issn}{00036951},
  \urlprefix\url{http://apl.aip.org/resource/1/applab/v63/i1/p36\_s1
  http://link.aip.org/link/APPLAB/v63/i1/p36/s1\&Agg=doi}.

\bibitem[{\citenamefont{Musil et~al.}(1990)\citenamefont{Musil, Kadlec,
  Valvoda, Ku\v{z}el, and \v{C}ern\'{y}}}]{Musil1990}
\bibinfo{author}{\bibfnamefont{J.}~\bibnamefont{Musil}},
  \bibinfo{author}{\bibfnamefont{S.}~\bibnamefont{Kadlec}},
  \bibinfo{author}{\bibfnamefont{V.}~\bibnamefont{Valvoda}},
  \bibinfo{author}{\bibfnamefont{R.}~\bibnamefont{Ku\v{z}el}},
  \bibnamefont{and}
  \bibinfo{author}{\bibfnamefont{R.}~\bibnamefont{\v{C}ern\'{y}}},
  \bibinfo{journal}{Surface and Coatings Technology}
  \textbf{\bibinfo{volume}{43-44}}, \bibinfo{pages}{259}
  (\bibinfo{year}{1990}), ISSN \bibinfo{issn}{02578972},
  \urlprefix\url{http://linkinghub.elsevier.com/retrieve/pii/025789729090079R}.

\bibitem[{\citenamefont{Poulek et~al.}(1989)\citenamefont{Poulek, Musil,
  \v{C}ern\'{y}, and Kuzel}}]{Poulek1989}
\bibinfo{author}{\bibfnamefont{V.}~\bibnamefont{Poulek}},
  \bibinfo{author}{\bibfnamefont{J.}~\bibnamefont{Musil}},
  \bibinfo{author}{\bibfnamefont{R.}~\bibnamefont{\v{C}ern\'{y}}},
  \bibnamefont{and} \bibinfo{author}{\bibfnamefont{R.}~\bibnamefont{Kuzel}},
  \bibinfo{journal}{Thin Solid Films} \textbf{\bibinfo{volume}{170}},
  \bibinfo{pages}{L55} (\bibinfo{year}{1989}), ISSN \bibinfo{issn}{00406090},
  \urlprefix\url{http://linkinghub.elsevier.com/retrieve/pii/0040609089907384}.

\bibitem[{\citenamefont{Poulek et~al.}(1991)\citenamefont{Poulek, Musil,
  Valvoda, and Kuzel}}]{Poulek1991}
\bibinfo{author}{\bibfnamefont{V.}~\bibnamefont{Poulek}},
  \bibinfo{author}{\bibfnamefont{J.}~\bibnamefont{Musil}},
  \bibinfo{author}{\bibfnamefont{V.}~\bibnamefont{Valvoda}}, \bibnamefont{and}
  \bibinfo{author}{\bibfnamefont{R.}~\bibnamefont{Kuzel}},
  \bibinfo{journal}{Thin Solid Films} \textbf{\bibinfo{volume}{196}},
  \bibinfo{pages}{265} (\bibinfo{year}{1991}), ISSN \bibinfo{issn}{00406090},
  \urlprefix\url{http://linkinghub.elsevier.com/retrieve/pii/004060909190370D}.

\bibitem[{\citenamefont{Grigorov et~al.}(1991)\citenamefont{Grigorov, Martev,
  Stoyanova, Vignes, and Langeron}}]{Grigorov1991}
\bibinfo{author}{\bibfnamefont{G.}~\bibnamefont{Grigorov}},
  \bibinfo{author}{\bibfnamefont{I.}~\bibnamefont{Martev}},
  \bibinfo{author}{\bibfnamefont{M.}~\bibnamefont{Stoyanova}},
  \bibinfo{author}{\bibfnamefont{J.-L.} \bibnamefont{Vignes}},
  \bibnamefont{and} \bibinfo{author}{\bibfnamefont{J.-P.}
  \bibnamefont{Langeron}}, \bibinfo{journal}{Thin Solid Films}
  \textbf{\bibinfo{volume}{198}}, \bibinfo{pages}{169} (\bibinfo{year}{1991}),
  ISSN \bibinfo{issn}{00406090},
  \urlprefix\url{http://linkinghub.elsevier.com/retrieve/pii/004060909190335U}.

\bibitem[{\citenamefont{Prenzel
  et~al.}(2013{\natexlab{a}})\citenamefont{Prenzel, Kortmann, von Keudell,
  Nahif, Schneider, Shihab, and Brinkmann}}]{Prenzel2013}
\bibinfo{author}{\bibfnamefont{M.}~\bibnamefont{Prenzel}},
  \bibinfo{author}{\bibfnamefont{A.}~\bibnamefont{Kortmann}},
  \bibinfo{author}{\bibfnamefont{A.}~\bibnamefont{von Keudell}},
  \bibinfo{author}{\bibfnamefont{F.}~\bibnamefont{Nahif}},
  \bibinfo{author}{\bibfnamefont{J.~M.} \bibnamefont{Schneider}},
  \bibinfo{author}{\bibfnamefont{M.}~\bibnamefont{Shihab}}, \bibnamefont{and}
  \bibinfo{author}{\bibfnamefont{R.~P.} \bibnamefont{Brinkmann}},
  \bibinfo{journal}{Journal of Physics D: Applied Physics}
  \textbf{\bibinfo{volume}{46}}, \bibinfo{pages}{084004}
  (\bibinfo{year}{2013}{\natexlab{a}}), ISSN \bibinfo{issn}{0022-3727},
  \urlprefix\url{http://stacks.iop.org/0022-3727/46/i=8/a=084004?key=crossref.3977466fb5377ee9197df775317007f8}.

\bibitem[{\citenamefont{Gautier et~al.}(1994)\citenamefont{Gautier, Fenaud,
  {Pham Van}, Villette, Pollak, Thromat, Jollet, and Duraud}}]{Gautier1994}
\bibinfo{author}{\bibfnamefont{M.}~\bibnamefont{Gautier}},
  \bibinfo{author}{\bibfnamefont{G.}~\bibnamefont{Fenaud}},
  \bibinfo{author}{\bibfnamefont{L.}~\bibnamefont{{Pham Van}}},
  \bibinfo{author}{\bibfnamefont{B.}~\bibnamefont{Villette}},
  \bibinfo{author}{\bibfnamefont{M.}~\bibnamefont{Pollak}},
  \bibinfo{author}{\bibfnamefont{N.}~\bibnamefont{Thromat}},
  \bibinfo{author}{\bibfnamefont{F.}~\bibnamefont{Jollet}}, \bibnamefont{and}
  \bibinfo{author}{\bibfnamefont{J.-p.} \bibnamefont{Duraud}},
  \bibinfo{journal}{Journal of the American Ceramic Society}
  \textbf{\bibinfo{volume}{77}}, \bibinfo{pages}{323} (\bibinfo{year}{1994}),
  ISSN \bibinfo{issn}{0002-7820},
  \urlprefix\url{http://doi.wiley.com/10.1111/j.1151-2916.1994.tb06999.x}.

\bibitem[{\citenamefont{Jiang et~al.}(2010)\citenamefont{Jiang, Sarakinos,
  Konstantinidis, and Schneider}}]{Jiang2010}
\bibinfo{author}{\bibfnamefont{K.}~\bibnamefont{Jiang}},
  \bibinfo{author}{\bibfnamefont{K.}~\bibnamefont{Sarakinos}},
  \bibinfo{author}{\bibfnamefont{S.}~\bibnamefont{Konstantinidis}},
  \bibnamefont{and} \bibinfo{author}{\bibfnamefont{J.~M.}
  \bibnamefont{Schneider}}, \bibinfo{journal}{Journal of Physics D: Applied
  Physics} \textbf{\bibinfo{volume}{43}}, \bibinfo{pages}{325202}
  (\bibinfo{year}{2010}), ISSN \bibinfo{issn}{0022-3727}.

\bibitem[{\citenamefont{Sarakinos et~al.}(2010)\citenamefont{Sarakinos, Music,
  Nahif, Jiang, Braun, Zilkens, and Schneider}}]{Sarakinos2010d}
\bibinfo{author}{\bibfnamefont{K.}~\bibnamefont{Sarakinos}},
  \bibinfo{author}{\bibfnamefont{D.}~\bibnamefont{Music}},
  \bibinfo{author}{\bibfnamefont{F.}~\bibnamefont{Nahif}},
  \bibinfo{author}{\bibfnamefont{K.}~\bibnamefont{Jiang}},
  \bibinfo{author}{\bibfnamefont{a.}~\bibnamefont{Braun}},
  \bibinfo{author}{\bibfnamefont{C.}~\bibnamefont{Zilkens}}, \bibnamefont{and}
  \bibinfo{author}{\bibfnamefont{J.~M.} \bibnamefont{Schneider}},
  \bibinfo{journal}{physica status solidi (RRL) - Rapid Research Letters}
  \textbf{\bibinfo{volume}{4}}, \bibinfo{pages}{154} (\bibinfo{year}{2010}),
  ISSN \bibinfo{issn}{18626254},
  \urlprefix\url{http://doi.wiley.com/10.1002/pssr.201004133}.

\bibitem[{\citenamefont{Music et~al.}(2011)\citenamefont{Music, Nahif,
  Sarakinos, Friederichsen, and Schneider}}]{Music2011a}
\bibinfo{author}{\bibfnamefont{D.}~\bibnamefont{Music}},
  \bibinfo{author}{\bibfnamefont{F.}~\bibnamefont{Nahif}},
  \bibinfo{author}{\bibfnamefont{K.}~\bibnamefont{Sarakinos}},
  \bibinfo{author}{\bibfnamefont{N.}~\bibnamefont{Friederichsen}},
  \bibnamefont{and} \bibinfo{author}{\bibfnamefont{J.~M.}
  \bibnamefont{Schneider}}, \bibinfo{journal}{Applied Physics Letters}
  \textbf{\bibinfo{volume}{98}}, \bibinfo{pages}{111908}
  (\bibinfo{year}{2011}), ISSN \bibinfo{issn}{00036951},
  \urlprefix\url{http://link.aip.org/link/APPLAB/v98/i11/p111908/s1\&Agg=doi}.

\bibitem[{\citenamefont{Rosén et~al.}(2006)\citenamefont{Rosén, Widenkvist,
  Larsson, Kreissig, Mráz, Martinez, Music, and Schneider}}]{Rosen2006}
\bibinfo{author}{\bibfnamefont{J.}~\bibnamefont{Rosén}},
  \bibinfo{author}{\bibfnamefont{E.}~\bibnamefont{Widenkvist}},
  \bibinfo{author}{\bibfnamefont{K.}~\bibnamefont{Larsson}},
  \bibinfo{author}{\bibfnamefont{U.}~\bibnamefont{Kreissig}},
  \bibinfo{author}{\bibfnamefont{S.}~\bibnamefont{Mráz}},
  \bibinfo{author}{\bibfnamefont{C.}~\bibnamefont{Martinez}},
  \bibinfo{author}{\bibfnamefont{D.}~\bibnamefont{Music}}, \bibnamefont{and}
  \bibinfo{author}{\bibfnamefont{J.~M.} \bibnamefont{Schneider}},
  \bibinfo{journal}{Applied Physics Letters} \textbf{\bibinfo{volume}{88}},
  \bibinfo{pages}{191905} (\bibinfo{year}{2006}), ISSN
  \bibinfo{issn}{00036951},
  \urlprefix\url{http://link.aip.org/link/APPLAB/v88/i19/p191905/s1\&Agg=doi}.

\bibitem[{\citenamefont{Schneider et~al.}(1999)\citenamefont{Schneider,
  Hjörvarsson, Wang, and Hultman}}]{Schneider1999}
\bibinfo{author}{\bibfnamefont{J.~M.} \bibnamefont{Schneider}},
  \bibinfo{author}{\bibfnamefont{B.}~\bibnamefont{Hjörvarsson}},
  \bibinfo{author}{\bibfnamefont{X.}~\bibnamefont{Wang}}, \bibnamefont{and}
  \bibinfo{author}{\bibfnamefont{L.}~\bibnamefont{Hultman}},
  \bibinfo{journal}{Applied Physics Letters} \textbf{\bibinfo{volume}{75}},
  \bibinfo{pages}{3476} (\bibinfo{year}{1999}), ISSN \bibinfo{issn}{00036951},
  \urlprefix\url{http://link.aip.org/link/APPLAB/v75/i22/p3476/s1\&Agg=doi}.

\bibitem[{\citenamefont{Kroesen et~al.}(1991)\citenamefont{Kroesen, Oehrlein,
  and Bestwick}}]{Kroesen1991}
\bibinfo{author}{\bibfnamefont{G.~M.~W.} \bibnamefont{Kroesen}},
  \bibinfo{author}{\bibfnamefont{G.~S.} \bibnamefont{Oehrlein}},
  \bibnamefont{and} \bibinfo{author}{\bibfnamefont{T.~D.}
  \bibnamefont{Bestwick}}, \bibinfo{journal}{Journal of Applied Physics}
  \textbf{\bibinfo{volume}{69}}, \bibinfo{pages}{3390} (\bibinfo{year}{1991}),
  ISSN \bibinfo{issn}{00218979},
  \urlprefix\url{http://link.aip.org/link/JAPIAU/v69/i5/p3390/s1\&Agg=doi}.

\bibitem[{\citenamefont{Shihab et~al.}(2012)\citenamefont{Shihab, Ziegler, and
  Brinkmann}}]{Shihab2012}
\bibinfo{author}{\bibfnamefont{M.}~\bibnamefont{Shihab}},
  \bibinfo{author}{\bibfnamefont{D.}~\bibnamefont{Ziegler}}, \bibnamefont{and}
  \bibinfo{author}{\bibfnamefont{R.~P.} \bibnamefont{Brinkmann}},
  \bibinfo{journal}{Journal of Physics D: Applied Physics}
  \textbf{\bibinfo{volume}{45}}, \bibinfo{pages}{185202}
  (\bibinfo{year}{2012}), ISSN \bibinfo{issn}{0022-3727},
  \urlprefix\url{http://stacks.iop.org/0022-3727/45/i=18/a=185202?key=crossref.32271c08a9a890ec429db0cee1751d5b}.

\bibitem[{\citenamefont{Levin and Brandon}(1998)}]{Brandon1998}
\bibinfo{author}{\bibfnamefont{I.}~\bibnamefont{Levin}} \bibnamefont{and}
  \bibinfo{author}{\bibfnamefont{D.}~\bibnamefont{Brandon}},
  \bibinfo{journal}{Journal of the American Ceramic Society}
  \textbf{\bibinfo{volume}{81}}, \bibinfo{pages}{1995} (\bibinfo{year}{1998}),
  ISSN \bibinfo{issn}{00027820},
  \urlprefix\url{http://dx.doi.org/10.1111/j.1151-2916.1998.tb02581.x
  http://doi.wiley.com/10.1111/j.1151-2916.1998.tb02581.x}.

\bibitem[{\citenamefont{Prenzel
  et~al.}(2013{\natexlab{b}})\citenamefont{Prenzel, Kortmann, von Keudell,
  Nahif, Schneider, Shihab, and Brinkmann}}]{Prenzel2012}
\bibinfo{author}{\bibfnamefont{M.}~\bibnamefont{Prenzel}},
  \bibinfo{author}{\bibfnamefont{A.}~\bibnamefont{Kortmann}},
  \bibinfo{author}{\bibfnamefont{A.}~\bibnamefont{von Keudell}},
  \bibinfo{author}{\bibfnamefont{F.}~\bibnamefont{Nahif}},
  \bibinfo{author}{\bibfnamefont{J.~M.} \bibnamefont{Schneider}},
  \bibinfo{author}{\bibfnamefont{M.}~\bibnamefont{Shihab}}, \bibnamefont{and}
  \bibinfo{author}{\bibfnamefont{R.~P.} \bibnamefont{Brinkmann}},
  \bibinfo{journal}{Journal of Physics D: Applied Physics}
  \textbf{\bibinfo{volume}{46}}, \bibinfo{pages}{084004}
  (\bibinfo{year}{2013}{\natexlab{b}}), ISSN \bibinfo{issn}{0022-3727},
  \urlprefix\url{http://stacks.iop.org/0022-3727/46/i=8/a=084004?key=crossref.3977466fb5377ee9197df775317007f8}.

\bibitem[{\citenamefont{Bruesch et~al.}(1984)\citenamefont{Bruesch, Koetz,
  Neff, and Pietronero}}]{Bruesch1984}
\bibinfo{author}{\bibfnamefont{P.}~\bibnamefont{Bruesch}},
  \bibinfo{author}{\bibfnamefont{R.}~\bibnamefont{Koetz}},
  \bibinfo{author}{\bibfnamefont{H.}~\bibnamefont{Neff}}, \bibnamefont{and}
  \bibinfo{author}{\bibfnamefont{L.}~\bibnamefont{Pietronero}},
  \bibinfo{journal}{Physical Review B} \textbf{\bibinfo{volume}{29}},
  \bibinfo{pages}{4691} (\bibinfo{year}{1984}).

\bibitem[{\citenamefont{Chu et~al.}(1988)\citenamefont{Chu, Bates, White, and
  Farlow}}]{Chu1988a}
\bibinfo{author}{\bibfnamefont{Y.~T.} \bibnamefont{Chu}},
  \bibinfo{author}{\bibfnamefont{J.~B.} \bibnamefont{Bates}},
  \bibinfo{author}{\bibfnamefont{C.~W.} \bibnamefont{White}}, \bibnamefont{and}
  \bibinfo{author}{\bibfnamefont{G.~C.} \bibnamefont{Farlow}},
  \bibinfo{journal}{Journal of Applied Physics} \textbf{\bibinfo{volume}{64}},
  \bibinfo{pages}{3727} (\bibinfo{year}{1988}), ISSN \bibinfo{issn}{00218979},
  \urlprefix\url{http://jap.aip.org/resource/1/japiau/v64/i7/p3727\_s1
  http://link.aip.org/link/JAPIAU/v64/i7/p3727/s1\&Agg=doi}.

\bibitem[{\citenamefont{M\"{o}ller and Eckstein}(1984)}]{TRIM2012}
\bibinfo{author}{\bibfnamefont{W.}~\bibnamefont{M\"{o}ller}} \bibnamefont{and}
  \bibinfo{author}{\bibfnamefont{W.}~\bibnamefont{Eckstein}},
  \bibinfo{journal}{Nuclear Instruments and Methods in Physics Research Section
  B: Beam Interactions with Materials and Atoms} \textbf{\bibinfo{volume}{2}},
  \bibinfo{pages}{814} (\bibinfo{year}{1984}), ISSN \bibinfo{issn}{0168583X},
  \urlprefix\url{http://linkinghub.elsevier.com/retrieve/pii/0168583X84903215}.

\end{thebibliography}

\end{document}